\documentstyle[12pt]{article}
\advance\textheight by \topskip
\begin{document}
 \begin{flushright}
SU-ITP-96-26\\
hep-th/9606093\\
June 16, 1996\\
\end{flushright}
\vspace{1 cm}
\begin{center}
\baselineskip=16pt

{\Large\bf    SUPERPOTENTIAL
\vskip .3cm

 FROM BLACK HOLES}

\vskip 2cm

{\bf Renata
Kallosh}\footnote { E-mail:
kallosh@physics.stanford.edu}\\
 \vskip 0.8cm
Physics Department, Stanford University, Stanford,   CA 94305-4060, USA\\
\vskip .6cm

\vskip 1 cm

\end{center}

\vskip 1 cm
\centerline{\bf ABSTRACT}
\begin{quotation}

 BPS monopoles in N=2 SUSY  theories  may lead to  monopole condensation and
confinement.  We have found that  supersymmetric  black holes with
non-vanishing area of the horizon may stabilize the moduli in theories where
the potential is proportional to the square of the graviphoton central charge.

In particular, in  known models of spontaneous breaking of N=2 to N=1 SUSY
theories, the parameters of the electric and magnetic Fayet--Iliopoulos terms
can be considered proportional to electric and magnetic charges of the
dyonic black holes. Upon such identification the potential  is found to be
proportional to the square of the black hole mass. The fixed values of the
moduli near the black hole horizon correspond exactly to the minimum of this
potential.  The value of the  potential at the minimum is proportional to the
black hole entropy.

 \end{quotation}

\newpage
 \baselineskip=15pt

N=2 supersymmetric  theories  may be viewed as low-energy effective actions
describing the non-perturbative dynamics of more fundamental theories. In
particular the vacuum structure and dyon spectrum of these theories has been
studied by Seiberg and Witten \cite{SW}  and the effect of BPS monopoles in N=2
supersymmetric Yang-Mills theories was understood. The relevance of generic
supersymmetric black holes to low-energy effective theories  was not studied
yet\footnote{Except for the  notable case of  black holes which could become
massless \cite{Smassless}.},  however, one may expect that such relation
exists. In particular, one could have  guessed that the black holes of N=2
theory with one half of supersymmetry unbroken may be somehow relevant to
models with spontaneous breaking of  N=2 supersymmetry    to N=1. We will find
indeed that this is the case and that the effect of supersymmetric dyonic black
holes is to stabilize the moduli. The choice of the superpotential  in such
models will be related to the central charge of the graviphoton, i.e. to the
black hole mass as the function of moduli and conserved charges of the black
hole.

The main difference between our study of the black holes  and the study of
non-perturbative phenomena in N=2 theories of rigid supersymmetries in
\cite{SW} is in the properties of the relevant central charge.

In ref. \cite{SW} the  central charge defining the mass of the dyon in rigid
supersymmetric theory is defined by charge of the  vector multiplet. It is
defined by  the symplectic section of a given N=2 theory $(X^\Lambda ,
F_\Lambda)$ and by conserved charges $(q_\Lambda , p^\lambda)$ of the  dyon and
  is given by
\begin{equation}
(M_{\rm YM}^{\rm  dyon})^2= |Z^{\rm rigid}(t, \bar t , q,p) |^2 =
\left |X^\Lambda(t)  q_\Lambda - F_\Lambda(t) \, p^\Lambda \right |^2 \ .
\label{rigidZ}\end{equation}
This formula defines the mass of the BPS dyons in supersymmetric Yang-Mills
theory.
In our case the central charge of the gravitational multiplet is the charge of
the graviphoton, supersymmetric  partner of the graviton in theories with local
supersymmetry.
It is defined  as follows \cite{Cer}:
\begin{equation}
(M_{\rm bh}^{\rm  dyon})^2  = |Z^{\rm local} (t, \bar t, q,p) |^2  = \left
| e^{K(t, \bar t)\over 2} [X^\Lambda(t)  q_\Lambda - F_\Lambda(t) \, p^\Lambda]
\right|^2 \ ,
\label{localZ}
\end{equation}
where $K(t, \bar t)$ is a Kahler potential
\begin{equation}
K = -\ln i( \bar X^\Lambda F_\Lambda -   X^ \Lambda  \bar F_\Lambda) \ .
\end{equation}
This formula for the values of the moduli $(t, \bar t)$ at infinity, far away
from the black hole, defines the ADM mass of the supersymmetric black hole
dyon.
 In all cases of supersymmetric black holes with non-vanishing area of the
horizon and
Bertotti-Robinson type geometry near the black hole horizon the following has
been proved \cite{FK}: The extremum of the square of the graviphoton central
charge (\ref{localZ}) in the moduli space relates the values of moduli to the
ratios of electric and magnetic charges. As a result, the value of the square
of the central charge at the extremum
in moduli space defines the area of the black hole horizon, which depends only
on conserved charges. The extremal value of this mass is also related to the
size of the infinite throat of the
Bertotti-Robinson geometry and is independent on the values of moduli far away
from the horizon:
\begin{equation}
\Bigl(|Z(t,\bar t (q,p) |^2 \Bigr)_{{\partial |Z(t,\bar t (q,p)  | \over
\partial t}=0} = {1\over \pi} \, S(q,p) \ .
\end{equation}

We will show in what follows that the idea of the supersymmetric attractors
\cite{FKS},  \cite{FK} which explains that  {\it moduli are driven to a fixed
point of attraction near the black hole horizon}, may be  realized in effective
field theories in a standard way: {\it scalars take the values  prescribed by
the minimum of the potential}. For this to happen we need 3 conditions to be
satisfied:

i)  to get the potential  depending on moduli $t,\bar t$  and some parameters
$(E,M)$ to be  proportional to the square of the central charge, depending on
the same moduli $t,\bar t$ and black hole charges  $(q,p)$.
\begin{equation}
V  \left ( t, \bar t, (E,M) \right) \sim   |Z^{\rm local} (t, \bar t, (q,p)
)|^2  =  \left
|  e^{K(t, \bar t)\over 2} [X^\Lambda(t)  q_\Lambda - F_\Lambda(t) \,
p^\Lambda] \right|^2 \ ,
\label{poten}\end{equation}

ii) the parameters $(E,M)$ in the potential have to be proportional to the
black hole charges
\begin{equation}
(E,M) \sim (q,p) \ ,
\label{relation}\end{equation}

iii) the relevant supersymmetric black hole has to have finite area of the
horizon for the potential to have a stable minimum with regular values of
moduli.

As we will show, these three conditions are satisfied in a specific model to be
discussed below. However
one may hope that they could be satisfied in more general situations.

It is important that the graviphoton central charge due to special properties
of local supersymmetry, which is doubled near the black hole horizon, has the
ability to stabilize the moduli in all cases of supersymmetric configurations
with finite area of the horizon. This follows from the universality of the
entropy-area formula for supersymmetric black holes proved in \cite{FK}. It is
not clear to us whether without black holes using only dyonic solutions of
rigid supersymmetric theories and the central charge formula (\ref{rigidZ}) one
can stabilize the moduli. At the same time various examples of stabilization of
moduli near the black hole horizon are known. Still no relation of any black
hole to superpotentials in low-energy theories has been established so far.

Spontaneous partial breaking of  N=2 SUSY to N=1  was discovered recently in
the context of globally supersymmetric theories by Antoniadis, Partouche and
Taylor  \cite{APT}. In what follows we will refer to this theory as  APT model.
The  partial supersymmetry breaking was also  found by a suitable flat limit of
local N=2 supergravity models by Ferrara, Girardello and Porrati \cite{FGP}, to
which we will refer as to FGP model. The APT-FGP potential has a stable
minimum, however  the  mechanism of partial spontaneous supersymmetry breaking
does not seem to have a natural dynamical explanation.

The new piece of information about black holes, to be presented below,  is  the
calculation of the attractor values of the
axion and dilaton for SL(2,Z) symmetric black holes \cite{KO}, which was not
performed before.  This  calculation triggered the identification  of the APT
potential and its minimum with  the axion-dilaton black hole near horizon,
giving us a first realization of the potential for moduli related to black
holes  as suggested in eq. (\ref{poten}).
The parameters $(E,M)$ in the potential turn out to be   the parameters of
electric and magnetic Fayet--Iliopoulos terms and they will be related to black
hole charges via the scale of supersymmetry breaking
\begin{equation}
(E,M) = \Lambda^2  (q,p) \ .
\end{equation}

We will start with the black hole side first. From all  known black holes it is
the family of axion-dilaton black holes with the manifest SL(2,Z) symmetry
\cite{KO}  which turns out to be relevant to the APT-FGP mechanism.
We will first consider the extreme axion-dilaton $U(1)\times U(1)$ black holes
of N=4 supergravity \cite{KO}. The corresponding black holes have been also
recognized as supersymmetric black holes of N=2 theory interacting with one
vector multiplet \cite{FKS}, \cite{KO}. There are two versions of this theory,
one corresponding to an SO(4) supergravity, which is described by the
prepotential $F= -i X^0 X^1$. The   other one, corresponding to the SU(4)
version of  N=4 theory  is related by a symplectic transformation to the first
one and has no prepotential but is defined by a symplectic section \cite{Cer}.
In all three versions of these theories one can study supersymmetric black
holes and their behavior near the horizon. Alternatively, one can simply find
the central charge matrix as a function of moduli and conserved quantized
charges and find the extremum of the largest eigenvalue of the central charge
\cite{FK}. This will define the fixed values of the moduli near the black hole
horizon as a function of charges.
\vskip 0.3cm

  N=4 {\it  axion-dilaton black holes}

 We will find the values of the dilaton and axion near the horizon directly
from the
$U(1)\times U(1)$
SL(2, Z) invariant black hole solutions \cite{KO} as   functions of conserved
electric and magnetic charges in two U(1) groups
$$n_1, \; m_1 \ , \qquad  n_2, \; m_2 \ .$$
The eigenvalues of the central charge matrix are given in \cite{KO}:
\begin{equation}
Z_1 = \sqrt 2(\Gamma_1 + i \Gamma_2)\ , \qquad Z_2 = \sqrt 2(\Gamma_1 - i
\Gamma_2) \ .
\label{centralcharge}\end{equation}
where
\begin{equation}
\Gamma_1 = {1\over 2} (Q_1 + i P_1)\ , \qquad \Gamma_1 = {1\over 2} (Q_1 + i
P_1) \ .
\label{gamma}\end{equation}
As found in eq. (30) of \cite{KO}  in terms of moduli and quantized charges we
have
\begin{eqnarray}
Q_1 + i P_1 &=& e^{\phi} (n_1 - \bar \lambda m_1) \ , \nonumber\\
 \nonumber\\
Q_2 + i P_2 &=& e^{\phi} (n_2 - \bar \lambda m_2) \ , \nonumber\\
\end{eqnarray}
where the axion-dilaton field is
\begin{equation}
\lambda = a + i e^{-\phi} \ .
\end{equation}
To find the values of $\lambda$ near the horizon we may use few methods. The
first is based on the fact that we know all solutions and we may simply pick up
the one in which moduli do not change and remain constant all the way from the
horizon to infinity. Such solution with unconstrained  four charges $n_1, \;
m_1 ,\;   n_2, \; m_2$
exists only in case that the moduli depend on these charges.  In our attractor
picture presented in Fig. 1 of \cite{FK} this corresponds to a solution which
is given by the horizontal line with zero slope. The  moduli and the metric for
the extreme  black hole solution is \cite{KO}
\begin{equation}
  \lambda (x) = {H_1(x) \over H_2(x)}
  \hskip 2 cm  ds^2 = (2{\rm Im} H_1  H_2)^{-1}  dt^2 -  2{\rm Im} H_1  H_2
(d\vec x )^2   \ ,
\end{equation}
where $H_1, H_2$ are harmonic functions. The solution under discussion is the
one with the vanishing axion-dilaton charge $\Upsilon$:
\begin{equation}
\bar \Upsilon = - { Z_1  Z_2 \over M} =0 \ .
\label{axdilcharge}\end{equation}

It follows from eqs. (\ref{centralcharge}), (\ref{gamma}), (\ref{axdilcharge})
that
\begin{equation}
  Q_1^1 + Q_2^2- P_1^2  - P_2^2  + i ( Q_1 P_1 + Q_2 P_2) =0 \ .
\label{fix}\end{equation}

\noindent The axion-dilaton field becomes

\begin{equation}
\lambda (r) = \left({ \frac{e^{\phi_{0}}}{\sqrt{2}}
                            \left[\lambda_{0}+
                            \frac{\lambda_{0}M+
                            \overline{\lambda}_{0}
                            \Upsilon }{r }\right] \over
\frac{e^{\phi_{0}}}{\sqrt{2}}
                            \left[1+\frac{M+
                            \Upsilon}{r
                            }\right]}\right) _{\Upsilon=0} =\  {\lambda_0 (1+
{M\over r}) \over  (1+ {M\over r})} = \lambda_0 \ ,
\end{equation}

\noindent where $ \lambda_0$ is the value of the moduli at infinity. For this
solution, however the value at infinity has to coincide with the value at the
horizon, which we may find by rewriting eqs. (\ref{fix}) in terms of moduli and
conserved charges:
\begin{equation}
e^{2 \phi}[ (n_1 -  \lambda m_1)^2  +  (n_2 - \lambda m_2)^2] =0 \ .
\end{equation}
This complex equation forces the constant complex moduli of this solution to
become a  function of charges:

\begin{equation}
(e^{-2\phi})_{\rm fix} = {|n_2 m_1 - n_1 m_2| \over m_1^2 + m_2^2}\ , \qquad
(a)_{\rm fix} = {n_2 m_2 + n_1 m_1 \over m_1^2 + m_2^2} \ .
\label{attr}\end{equation}

We may also verify that according to \cite{FK} the same values of moduli at the
fixed point may be obtained by the extremizing of the square of the largest
eigenvalue of the central charge.
\begin{equation}
|Z_1 |^2= 2 |\Gamma_1 + i \Gamma_2|^2 = {1\over 2} | e^{\phi} (n_1 - \bar
\lambda m_1) +i
e^{\phi} (n_2 - \bar \lambda m_2)|^2 \ .
\label{centr}\end{equation}
It can be checked that the  equations for the extremum of the central charge
\begin{equation}
 {\partial |Z(\lambda ,  \bar \lambda (n_1, m_1, n_2,m_2)  |^2 \over \partial
\lambda } =0
\end{equation}
are solved by the fixed point values of moduli in eqs. (\ref{attr}).
We may also note that the value of the central charge at the extremum is
proportional to the black hole entropy, which for these black holes was found
in \cite{KO}, \cite{FK}  to be equal to

\begin{equation}
S=  \pi (|Z_1|^2 -  |Z_2|^2 ) =\pi (|Z_1|^2 )_{\rm fix} = 2 \pi |n_1 m_2 - n_2
m_1| \ .
\end{equation}

We will proceed with identification of the corresponding values of moduli near
the black hole horizon in the two versions of N=2 theory.  The details will be
presented in a separate publication \cite{KW}.
\vskip 0.3 cm

N=2 {\it black holes in the theory with the prepotential} $F = -i X^0 X^1$
(SO(4) {\it version of } N=4)

 In the version with the prepotential we have
$$q_0, \; p^0\ , \qquad  q_1, \; p^1 \ , $$
and the moduli take the values near the horizon
\begin{equation}
(e^{-2\phi})_{\rm fix} = {|q_0 q_1 +  p^0 p^1| \over (q_1)^2 + (p^0)^2} \qquad
(a)_{\rm fix} = {p^1 q_1 -  q_0 p^0  \over (q_1)^2 + (p^0)^2} \ .
\end{equation}
\vskip 0.3 cm

N=2 {\it black holes in the theory without prepotential } (SU(4) {\it version
of } N=4)

In the version which has no prepotential we have the charges
$$\hat q_0, \; \hat p^0 \qquad  \hat q_1, \; \hat p^1 \ ,$$
and the moduli are
\begin{equation}
(e^{-2\hat \phi})_{\rm fix} = {|\hat q_1 \hat p^0  -  \hat q_0|  \hat  p^1
\over (\hat p^1)^2 + (\hat p^0)^2}\ , \qquad
(a)_{\rm fix} = {- \hat q_1 \hat p^1  -  \hat q_0 \hat p^0  \over (\hat p^1)^2
+ (\hat p^0)^2} \ .
\end{equation}
For all these cases it is easy to   identify the relation between charges of
these three versions of the same theory as well as central charges.
\vskip 0.3 cm

{\it Superpotential in APT-FGP model and black holes near the horizon}

APT model as well as flat limit of FGP model\footnote{We use the notation for
the F-I parameters
as given in FGP model and we consider  $b^0$,   the constant value of one of
the quaternionic coordinates of the hypermultiplet manifold, to be equal to 1.}
 in terms of manifest  N=1 supersymmetry consist of   usual terms
\begin{equation}
-{ i\over 4}\int d^2 \theta \lambda \, {\cal W} + c.c. + \int d^2 \theta  d^2
\bar \theta K \ ,
\end{equation}
where ${\cal W}$ is the gauge field superfield and $K$ is the Kahler potential.
 This action is supplemented by the  F-I term
\begin{equation}
\Lambda^2 \sqrt 2 \xi D \  ,
\end{equation}
as well as by an unusual superpotential term
\begin{equation}
\Lambda^2 \int d^2 \theta W + c.c. \ .
\end{equation}
Here
\begin{equation}
W= eb + m{\cal F}_b
\end{equation}
and $b$ is a chiral superfield and ${\cal F}_b$ is the derivative of the
prepotential over $b$. In terms of the manifest N=2 superfields the APT
Lagrangian is
\begin{equation}
{i\over 4} \int d^2 \theta_1  d^2  \theta_2 \, [ {\cal F}(B) - B^D B] + {1\over
2}(\vec E\cdot \vec Y + \vec M\cdot \vec Y^D) + c.c. \ ,
\end{equation}
where $B$ is an unconstrained chiral N=2 multiplet,  $B^D$ is a  constrained
chiral N=2 multiplet  playing the role of the Lagrange multiplier.  The
constant vectors
 $\vec E, \vec M$  ($\vec M$ being real)  define electric and  magnetic
Fayet--Iliopoulos terms, since they are the coefficients in front of  the
auxiliary fields of the $B, B^D$ multiplets, $\vec Y , \vec Y^D$. The  N=2
auxiliary fields   form  SU(2) triplets since they combine the complex
auxiliary field of N=1 chiral multiplet  $F+iG$ with
the auxiliary field of N=1 vector multiplet $D$.
 In APT-FGP model the F-I parameters which lead to spontaneous breaking of N=2
supersymmetry to N=1
are chosen to be
\begin{equation}
{\rm Re} \vec E = \Lambda^2 (0,e,\xi)\ , \qquad  \vec M = \Lambda^2 (0,m, 0) \
{}.
\end{equation}
Upon elimination of auxiliary fields the scalar potential   is
\begin{equation}
V= { | {\rm Re}\vec E + \lambda  \vec M|^2 \over {\rm Im} \lambda} + \dots \ ,
\label{APTpot}\end{equation}
where $\dots$ denote terms independent of moduli $\lambda$. These terms  differ
in APT
model  and in the flat limit of  FGP model.  At the moment from the black hole
side we do not have information on field independent part of the potential and
in what follows we will compare
the APT-FGP field dependent part of the potential with the black hole central
charge.
A stable minimum of the potential for the scalar fields in APT theory exists at
\begin{equation}
({\rm Im} \lambda)_{\rm min} = (e^{-2\phi})_{\rm min} =\left| {\xi  \over
m}\right| \ , \qquad
({\rm Re} \lambda )_{\rm min}= (a)_{\rm min} = -{  e \over m} \ .
\label{min}\end{equation}

It is fairly easy to see that if one would take any of the 3 types of black
hole solutions above
and in each case choose only 3 non-vanishing charges one would always reproduce
the relevant values of the scalars at the minimum of the potential from the
fixed points of moduli in black holes. In particular, in the first case we may
take a black hole with
\begin{equation}
m_2=0 \ , \qquad (e^{-2\phi})_{\rm fix}= \left|{n_2\over m_1}\right|
=\left|{\xi  \over m}\right| =(e^{-2\phi})_{\rm min}\ ,  \qquad  (a)_{\rm fix}=
{n_1\over m_1}=- { e  \over m}=  (a)_{\rm min} \ .
\label{N=4}\end{equation}
In the second case
\begin{equation}
p^0=0\ , \qquad (e^{-2\phi})_{\rm fix}= \left|{q_0\over q_1}\right|=\left|{\xi
\over m}\right| =(e^{-2\phi})_{\rm min}\ ,  \qquad  (a)_{\rm fix}= {p^1\over
q_1}=-{ e  \over m}=  (a)_{\rm min} \ .
\label{prepot}\end{equation}
In the third case we need
\begin{equation}
\hat p^1=0\ , \qquad (e^{-2\phi})_{\rm fix}= \left|{\hat q_1\over \hat
p^1}\right|=\left|{\xi  \over m}\right| =(e^{-2\phi})_{\rm min} \ , \qquad
(a) _{\rm fix}= {\hat q_0 \over \hat p^1}=-{e  \over m}=  (a)_{\rm min} \ .
\label{su4}\end{equation}
This  gives the relation between the ratios of parameters of F-I terms leading
to spontaneous breaking of N=2 supersymmetry to N=1 supersymmetry in APT-FGP
model and  ratios of charges of supersymmetric black holes with 1/4 of unbroken
N=4 supersymmetry or 1/2 of N=2 one.

Let us now compare the APT-FGP  potential (\ref{APTpot}) with the square of the
graviphoton central charge. The field dependent part of the potential is
\begin{equation}
V \sim   e^{2 \phi} \left[ (e+ ma) ^2  +  (e^{- \phi} m   +  \xi) ^2  \right] +
\dots \ ,
\label{potential}\end{equation}
and its value at infinity is
\begin{equation}
V_{\rm min}  \sim   2| m\xi| \ .
\label{potentialmin}\end{equation}

We  take the N=4 black hole central charge (\ref{centr}) at $m_2=0$, i.e.
restricting the 4-charge solution to only 3-charge solution. We get
\begin{equation}
|Z |^2 = {1\over 2}  e^{2 \phi} \left [(n_1-am_1)^2 + (m_1 e^{- \phi} +n_2)^2
\right ] \ ,
\label{centr2}\end{equation}
and its value at the fixed point is
\begin{equation}
|Z |^2_{\rm fix} = 2 |m_1 n_2|  \ .
\label{centrfix}\end{equation}

Relation (\ref{N=4})  between F-I parameters of the APT-FGP model and black
hole charges,
$$\left|{n_2\over m_1}\right| =\left|{\xi  \over m}\right| , \qquad {n_1\over
m_1}=-{e  \over m} \ ,$$
which was found before
 from the minimal values of the potential and from the fixed point  of the
values of the moduli near black hole horizon is confirmed. The same can be
established for other cases. With such identification the potential coincides
with the black hole central charge as a function of moduli in the generic point
of the moduli space.
It is quite remarkable that the procedure of variation of the central charge
over the moduli at fixed conserved charges of the black hole  suggested in
\cite{FK} becomes the procedure of variation of the potential over the scalar
fields at the fixed values of the F-I parameters. The minimum of the field
dependent part of the potential is proportional to the minimum of the central
charge, i.e. proportional to the black hole entropy.
\begin{equation}
V_{\rm min} (E,M) \sim \pi (|Z  (q,p)|^2)_{\rm fix} = S(q,p) \ .
\end{equation}

 In  conclusion,   we have found  an interesting correspondence between
specific Lorentz-covariant  models of   N=2 supersymmetry spontaneously broken
down  to N=1   \cite{APT,FGP} and  the axion-dilaton black holes: the field
dependent part of the potential is proportional to the square of the black hole
central charge, considered in the generic point of moduli space. The minimum of
the potential is the minimum of the central charge extremized in the moduli
space. According to  \cite{FK} this minimum describes the stabilization  of the
moduli near the black hole horizon. The  APT-FGP model clearly shows that the
origin of the superpotential stabilizing the moduli is related to
supersymmetric black holes with non-vanishing area of the horizon. It is also
important to stress that whereas black holes require the concept of static
geometry breaking Lorentz invariance,  the   low-energy action with partially
spontaneously broken supersymmetry describes Lorentz-covariant theory which
codifies the most important properties of extreme black holes with partially
broken supersymmetry. In particular, the value of the potential at the minimum
is proportional to the black hole entropy.
In  \cite{SW} the effects of  non-gravitational dyons (solitons of gauge
theories) on supersymmetric low-energy theories were studied. In our case of
gravitational dyons (black holes) a new phenomenon  which was not seen before
is the
 stabilization of the moduli as a generic property of black holes with
non-vanishing entropy.

Our observation opens various directions for future investigation.  During the
last few years  many new black hole solutions which break  1/2, 3/4 or 7/8 of
the original supersymmetry have been found. It would be very interesting to
find the counterparts of   these solutions in supersymmetric models of
elementary particles.  It seems likely that new  models of global supersymmetry
 may be constructed which are related to   generic supersymmetric black holes
in such a way that the potential is proportional to the square of the black
hole central charge.   Such models   will have build-in duality symmetry, since
the black hole mass formulae have  such symmetry.   It may be possible, in
particular, to construct models with global N=4 supersymmetry with the black
hole type potential for breaking the supersymmetry spontaneously down to N=1,
since there exist   black holes which   break 3/4 of N = 4 supersymmetry.   One
may try to use the fact that  theories with global N=4 supersymmetry and some
N=2 theories are finite and therefore the new models  of N=1 supersymmetry have
to be finite, since the breaking of supersymmetry from N=4 and from N=2 to N=1
in such theories  will be  spontaneous.

It would be interesting to  combine  the picture discussed above with
hypermultiplets to include massless black holes  and to find  the relation of
such models to string theory and their vacua.   It would be especially
important  to understand the dynamical mechanism behind the relation between
the potential leading to partial spontaneous breaking
of supersymmetry and the black hole mass formula.

\section{Acknowledgements}
 Stimulating discussions with   A. Linde and L. Susskind  are gratefully
acknowledged.
The work  is supported by the  NSF grant
PHY-9219345.

\end{document}